\documentclass{article}

\begin{document}

\centerline{{\LARGE No-cloning with unitary scaling}}

\centerline{Dafa Li}
\centerline{Department of Mathematical Sciences, Tsinghua
University, Beijing 100084, CHINA} \centerline{email: lidafa@tsinghua.edu.cn}
\centerline{Corresponding author: Dafa Li}

Abstract. It is known that the classical information like strings of bits
can be copied. In 1982, Wootters and Zurek proposed the quantum no-cloning
principle \cite{Wootters}. No-cloning principle says that it is impossible
to make an identical copy of an arbitrary unknown quantum pure state by
using unitary evolution. In this paper, we call $\mathcal{U}|\psi \rangle $,
where $\mathcal{U}$ is any unitary operator, a $\mathcal{U-}$copy of the
state $|\psi \rangle $ and show  it is impossible to make a $\mathcal{U-}$%
copy of an arbitrary unknown quantum pure state by using unitary evolution.

\section{Introduction}

The quantum no-cloning theorem, formulated in 1982, is a fundamental
principle which is different from classic information \cite{Wootters}. It is
known that the classical information can be copied. No-cloning theorem says
that it is impossible to make an identical copy of an arbitrary unknown
quantum pure state by using unitary evolution. This constraint arises from
the linearity of quantum mechanics. No-cloning theorem does not prevent all
quantum states being copied. Two orthogonal states can be perfectly copied
\cite{Nielsen}. For example, two states $\frac{1}{\sqrt{2}}(|0\rangle \pm
|1\rangle )$ are orthogonal and hence, they can be perfectly\ copied.
No-cloning theorem only confirms two non-orthogonal states cannot be
perfectly copied. For example, two states $\frac{1}{\sqrt{2}}(|0\rangle
+|1\rangle )$ and $\sqrt{\frac{1}{3}}|0\rangle +\sqrt{\frac{2}{3}}|1\rangle )
$ are not orthogonal, so they cannot be perfectly copied. Duan and Guo
proposed probabilistic quantum cloning machine. The cloning\ machine can
obtain perfect copies with a probability less than 1 \cite{Duan}.

No-cloning theorem yields the significant effects in quantum computing,
quantum teleportation, quantum cryptography and quantum key distribution. In
classical computation, errors can be corrected by using the redundancy
obtained by duplicating bits. In quantum communication and computation, the
quantum no-cloning theorem makes error correction impossible using the
redundancy. Bennett et al. and Shor proposed quantum error correction coding
\cite{Bennett, Shor}.

When a quantum state is transferred, the original state and the resource
state for teleportation are rewritten, after measuring then the original
state is destroyed. Clearly, the existence of two copies cannot happen.
No-cloning theorem is foundational for quantum cryptography and security. It
ensures secure, eavesdropper-detecting communication.

No-hiding theorem and no-broadcasting were proposed in \cite{Samuel} and
\cite{Barnum}, respectively. No-deleting theorem says there is no quantum
deleting machine which can delete one unknown state against a copy \cite%
{Pati}.

\section{No-cloning with unitary scaling}

In \cite{Wootters, Nielsen}, the cloning machine is defined as

\begin{equation}
T|\psi \rangle |0\rangle =|\psi \rangle |\psi \rangle .  \label{cp-1}
\end{equation}%
\ \ \ The machine starts out in an unknown pure state $|\psi \rangle $. The
initial state of the machine is $|\psi \rangle |0\rangle $, where $|0\rangle
$ is the ready state of the target system. By a unitary evolution $T$, the
cloning machine makes an identical copy of an arbitrary unknown quantum pure
state $|\psi \rangle $. The cloning machine in Eq. (\ref{cp-1})\ is called a
standard cloning machine.

A copy machine at an office with color, black-and-white, and scaling
capabilities allows users to adjust document sizes and colors via the
control panel. In this paper, we explore that if there is a quantum cloning
machine which can make a non-identical copy of an arbitrary unknown quantum
pure state.

The cloning machine with unitary scaling is defined for an input state $%
|\psi \rangle |0\rangle $ as follows:

\begin{equation}
T|\psi \rangle |0\rangle =|\psi \rangle \mathcal{U}|\psi \rangle ,
\label{cp-2}
\end{equation}%
where $T$\ is\ a unitary evolution, $|\psi \rangle $ is an arbitrary unknown
quantum pure state, $|0\rangle $ is the ready state of the target system, $%
\mathcal{U}$ is any unitary operator, and $\mathcal{U}|\psi \rangle $ is
called a $\mathcal{U}$-copy of $|\psi \rangle $. The cloning machine with
unitary scaling makes a $\mathcal{U}$-copy of an arbitrary unknown quantum
pure state $|\psi \rangle $.

When $\mathcal{U}$ is the identity, the cloning machine\ with unitary
scaling in Eq. (\ref{cp-2}) reduces to the standard cloning machine in Eq. (%
\ref{cp-1}).

For example, let $|\psi \rangle =\alpha |0\rangle +\beta |1\rangle $ and $H$
be the Hadmard operator. We try to make a $H$-copy of the unknown state $%
|\psi \rangle $ by using a CNOT gate as follows. Let $T$ be the CNOT gate.\
From Eq. (\ref{cp-2}), we have the following
\begin{equation}
T|\psi \rangle |0\rangle =|\psi \rangle \mathcal{H}|\psi \rangle .
\end{equation}%
Then, a calculation yields
\begin{equation}
|\psi \rangle \mathcal{H}|\psi \rangle =\frac{\alpha (\alpha +\beta )}{\sqrt{%
2}}|00\rangle +\frac{\alpha (\alpha -\beta )}{\sqrt{2}}|01\rangle +\frac{%
\beta (\alpha +\beta )}{\sqrt{2}}|10\rangle +\frac{\beta (\alpha -\beta )}{%
\sqrt{2}}|11\rangle ,  \label{ex-2}
\end{equation}%
\begin{eqnarray}
|\psi \rangle |0\rangle  &=&\alpha |00\rangle +\beta |10\rangle ,
\label{ex-3} \\
T|\psi \rangle |0\rangle  &=&\alpha |00\rangle +\beta |11\rangle .
\label{ex-4}
\end{eqnarray}

From Eqs. (\ref{ex-2}, \ref{ex-4}), one can see that $T|\psi \rangle
|0\rangle =|\psi \rangle \mathcal{H}|\psi \rangle $\ unless $\alpha =\beta
=0 $. It means that it is impossible to make a H-copy of an arbitrary
unknown quantum pure state by using CNOT.

We next argue that it is impossible to make a $\mathcal{U}$-copy of an
arbitrary unknown quantum pure state by using unitary evolution. Assume that
the cloning machine with unitary scaling works for pure states $|\psi
\rangle $ and $|\phi \rangle $. Then, we obtain
\begin{eqnarray}
T|\psi \rangle |0\rangle  &=&|\psi \rangle \mathcal{U}|\psi \rangle ,
\label{equa-1} \\
T|\phi \rangle |0\rangle  &=&|\phi \rangle \mathcal{U}|\phi \rangle .
\label{equa-2}
\end{eqnarray}

A calculation yields the inner product of the above two equations below.

\begin{equation}
\langle \psi |\langle 0|T^{H}T|\phi \rangle |0\rangle =\langle \psi |\langle
\psi |\mathcal{U}^{H}|\phi \rangle (\mathcal{U}|\phi \rangle ).  \label{eq-5}
\end{equation}%
Note that $T$ is a unitary evolution and $T^{H}$\ is the Hermitian transpose
of $T$. From the LHS of Eq. (\ref{eq-5}), obtain
\begin{equation}
\langle \psi |\langle 0|T^{H}T|\phi \rangle |0\rangle =\langle \psi |\langle
0||\phi \rangle |0\rangle =\langle \psi |\phi \rangle .  \label{eq-6}
\end{equation}%
From the RHS of Eq. (\ref{eq-5}), obtain
\begin{equation}
\langle \psi |\langle \psi |\mathcal{U}^{H}|\phi \rangle (\mathcal{U}|\phi
\rangle )=\langle \psi ||\phi \rangle \langle \psi |\mathcal{U}^{H}\mathcal{U%
}|\phi \rangle =(\langle \psi |\phi \rangle )^{2}.  \label{eq-7}
\end{equation}%
From Eqs. (\ref{eq-5}, \ref{eq-6}, \ref{eq-7}), obtain
\begin{equation}
\langle \psi |\phi \rangle =(\langle \psi |\phi \rangle )^{2}.  \label{q-8}
\end{equation}%
Thus, $\langle \psi ||\phi \rangle =0$, or $\langle \psi ||\phi \rangle =1$.
For the case $\langle \psi ||\phi \rangle =0$, $|\psi \rangle $ and $|\phi
\rangle $ are orthogonal. For the case $\langle \psi ||\phi \rangle =1$, $%
|\psi \rangle =|\phi \rangle $. Thus, the cloning machine with unitary
scaling works only for two mutually orthogonal states but not for two
non-orthogonal states, and therefore, the cloning machine with unitary
scaling  cannot make a $\mathcal{U}$-copy of an arbitrary unknown quantum
pure state.

We can next show that the cloning machine\ with unitary scaling\ does not
work for a superposition $p|\psi \rangle +q|\phi \rangle $, where $pq\neq 0$%
, of two orthogonal states $|\psi \rangle $ and $|\phi \rangle $. For $|\psi
\rangle $ and $|\phi \rangle $, we obtain Eqs. (\ref{equa-1}, \ref{equa-2}).
Let
\begin{equation}
|\omega \rangle =p|\psi \rangle +q|\phi \rangle .  \label{q-10}
\end{equation}%
Assume that the cloning machine\ with unitary scaling can work for $|\omega
\rangle $, i.e.
\begin{equation}
T|\omega \rangle |0\rangle =|\omega \rangle \mathcal{U}|\omega \rangle .
\label{q-11}
\end{equation}%
From Eqs. (\ref{equa-1}, \ref{equa-2}), the LHS\ of Eq. (\ref{q-11}) becomes
\begin{equation}
T|\omega \rangle |0\rangle =pT|\psi \rangle |0\rangle +qT|\phi \rangle
|0\rangle =p|\psi \rangle \mathcal{U}|\psi \rangle +q|\phi \rangle \mathcal{U%
}|\phi \rangle .  \label{q-12}
\end{equation}%
Via Eq. (\ref{q-10}), the RHS\ of Eq. (\ref{q-11}) becomes
\begin{equation}
|\omega \rangle \mathcal{U}|\omega \rangle =p^{2}|\psi \rangle \mathcal{U}%
|\psi \rangle +pq|\psi \rangle \mathcal{U}|\phi \rangle +pq|\phi \rangle
\mathcal{U}|\psi \rangle +q^{2}|\phi \rangle \mathcal{U}|\phi \rangle .
\label{q-13}
\end{equation}%
Clearly, from Eqs. (\ref{q-12}, \ref{q-13}), Eq. (\ref{q-11}) holds unless $%
pq=0$.

\section{The relation between no-cloning and no-cloning with unitary scaling
}

If the standard cloning machine can make an identical copy of an arbitrary
unknown pure state $|\psi \rangle $, i.e.
\begin{equation}
T|\psi \rangle |0\rangle =|\psi \rangle |\psi \rangle ,
\end{equation}%
then for any unitary operator $\mathcal{U}$,
\begin{equation}
(I\otimes \mathcal{U)}T|\psi \rangle |0\rangle =(I\otimes \mathcal{U)}|\psi
\rangle |\psi \rangle =|\psi \rangle \mathcal{U}|\psi \rangle .
\end{equation}%
Clearly, $(I\otimes \mathcal{U)}T$ is unitary. It means that $(I\otimes
\mathcal{U)}T$ can make a $\mathcal{U}$-copy of $|\psi \rangle .$

Conversely, if $T$ can make a $\mathcal{U}$-copy of an arbitrary unknown
pure state$\ |\psi \rangle $, i.e.
\begin{equation}
T|\psi \rangle |0\rangle =|\psi \rangle \mathcal{U}|\psi \rangle ,
\end{equation}%
then
\begin{equation}
(I\otimes \mathcal{U}^{H}\mathcal{)}T|\psi \rangle |0\rangle =(I\otimes
\mathcal{U}^{H}\mathcal{)}|\psi \rangle \mathcal{U}|\psi \rangle =|\psi
\rangle |\psi \rangle .
\end{equation}%
Clearly, $(I\otimes \mathcal{U}^{H}\mathcal{)}T$ is also unitary. Thus, $%
(I\otimes \mathcal{U}^{H}\mathcal{)}T$ can make an identical copy of the
unknown state$\ |\psi \rangle $.

Therefore, the standard cloning machine can make an identical copy of an
arbitrary unknown pure state if and only if the cloning machine\ with
unitary scaling can make a $\mathcal{U}$-copy of the state.

\end{document}